\documentclass[conference]{IEEEtran}
\usepackage{cite}
\IEEEoverridecommandlockouts
\usepackage{booktabs} 
\usepackage{graphicx}
\usepackage{enumitem}
\usepackage{paralist}
\usepackage{listings}
\usepackage{color}
\usepackage{caption}
\usepackage{algorithm}
\usepackage[noend]{algpseudocode}
\usepackage{subfig}
\usepackage[utf8]{inputenc}
\usepackage{url}
\usepackage{float}
\usepackage{enumitem}
\usepackage{svg}
\usepackage{comment}
\usepackage{xspace}

\definecolor{dkgreen}{rgb}{0,0.6,0}
\definecolor{gray}{rgb}{0.5,0.5,0.5}
\definecolor{mauve}{rgb}{0.58,0,0.82}

\newcommand{\admnote}[1]{{\textcolor{blue}{ADM: #1 }}\xspace}

\begin{document}

\bstctlcite{IEEEexample:BSTcontrol}

\title{The Ghost of Performance Reproducibility Past}


\author{%
\IEEEauthorblockN{Srinivasan Ramesh\IEEEauthorrefmark{1}
        Mikhail Titov\IEEEauthorrefmark{3}
        Matteo Turilli\IEEEauthorrefmark{2}\IEEEauthorrefmark{3}
        Shantenu Jha\IEEEauthorrefmark{2}\IEEEauthorrefmark{3}
        Allen Malony\IEEEauthorrefmark{1}
}
\IEEEauthorblockA{\IEEEauthorrefmark{1}University of Oregon
        \{sramesh, malony\}@cs.uoregon.edu%
}

\IEEEauthorblockA{\IEEEauthorrefmark{2}Rutgers University, New Brunswick}

\IEEEauthorblockA{\IEEEauthorrefmark{3}Brookhaven National Laboratory
        \{mtitov, mturilli, shantenu\}@bnl.gov%
}

}

\maketitle
\begin{abstract}

The importance of ensemble computing is well established. However, executing
ensembles at scale introduces interesting performance fluctuations that have
not been well investigated. In this
paper, we trace our experience
uncovering performance fluctuations of ensemble applications
(primarily constituting a workflow of GROMACS tasks),
and unsuccessful attempts, so far, at trying to discern the
underlying cause(s) of performance fluctuations. Is the failure to discern the
causative or contributing factors a failure of capability? Or imagination? Do
the fluctuations have their genesis in some inscrutable aspect of the
system or software? Does it warrant a fundamental reassessment
and rethinking of how we assume and conceptualize performance reproducibility?
Answers to these questions are not straightforward, nor are they immediate or
obvious. We conclude with a discussion about the performance of ensemble
applications and ruminate over the implications for how we define and measure 
application performance.




\end{abstract}




\maketitle

\section{Introduction}

Scientific high-performance computing (HPC) has focused on a workload's functionality,
scale, and performance with a single large task. However, the end
of Dennard scaling, coupled with the realities of post-Moore parallelism,
places increasingly severe limitations on the ability of a single monolithic
task to achieve significant performance gains on large-scale parallel
machines. In response to these challenges, scientific workflows have emerged
as an important way of developing scientific applications on HPC platforms~\cite{jha2019incorporating}.
The coupling of AI to traditional HPC simulations is further increasing the performance and
the sophistication of scientific workflow
applications~\cite{lee2020scalable,brace2022coupling}.


Ensemble computing represents a particular case of workflows in which multiple
copies of the same application are executed concurrently but independently.
Ensemble-based applications have been used to generate improved scientific
insight in multiple
domains~\cite{balasubramanian2018harnessing,brace2022coupling,cosb18kasson}.
Today, ensemble-based applications executing on HPC platforms orchestrate 100s
to 1000s of, individual tasks. Typically, each task is a scientific simulation
implemented as a parallel MPI program spanning one or more distributed
computing nodes. Further, the tasks can be heterogeneous regarding the
application they represent and the computing resources to which they are
mapped. Applications on the horizon~\cite{balasubramanian2020adaptive} will
require the concurrent execution of $10^6-10^8$ MPI tasks,  the implications
of which are many-fold and profound~\cite{merzky2022raptor}

This work investigates the issue of the performance of an ensemble of tasks.
In particular, it explores variability in the performance (as measured by the
time-to-execution) of a set of otherwise identical tasks executed
concurrently. Contrary to expectation, individual tasks did not have identical
times-to-execution. The variability in runtime is seemingly independent of the
specific executable and time-invariant (viz., state-of-the HPC machine). This paper
builds upon and extends the result earlier identified
~\cite{pouchard2019computational,pouchard2018use}.
Whether the execution time of otherwise identical tasks is indeed scale and time
invariant, and independent of the specific task remain
open questions.




To address these open questions, we investigate the performance of an ensemble
of tasks, and measure the performance of an individual task taken from a set
of otherwise identical tasks. We also evaluate the performance of the same set
of tasks (workload) run at different instances of time. Lastly, we investigate
the performance variability as a function of the workload size (i.e., the
number of tasks).

The key contributions of this experience paper are:
\begin{inparaenum}
  \item Demonstrating and investigating performance variability of individual
  tasks that are part of a set of concurrently executing tasks;
  \item Characterizing spatio-temporal aspects of the performance variability;
  \item Design of experiments to investigate and narrow the possible causes of performance variability;
  \item Postulating possible causes of performance variability and challenges in establishing performance reproducibility; and
  \item Synthesizing the insight gained towards a nuanced yet generalized
  perspective of performance and performance reproducibility.
\end{inparaenum}

\section{Characterizing Performance Variation in GROMACS Ensembles}

Homogeneous ensemble execution typically consists of simultaneously executing
copies of the same program (ensemble task) on different input decks,
representing for example, different initial conditions or input execution
parameters. When input decks are different, individual ensemble tasks may
execute different code paths leading to some natural variation in the
individual task runtime. However, when the ensemble tasks execute
\textit{identical} input decks on \textit{identical} resource allocations, any
variation in task performance must result from system (exogenous) factors.
Examples of these system factors known to cause application performance
variation include processor variability~\cite{PROCESSOR}, inter-job network
interference~\cite{NETWORK}, and operating system (OS) noise~\cite{CHUNDURI}.


Experiments in Section II assumes that all tasks in the workload
\textit{exactly} concurrently fit in the allocated node resources. In such a scenario,
\textit{makespan} or the total execution time of the workload is affected by
the task runtime variation arising from system (exogenous) factors. Two
questions arise in this scenario:
\begin{inparaenum}[Q-1]
    \item  What is the range of the execution times of individual ensemble
    tasks? Equally, how much do system factors affect the individual task
    runtime?
    \item  What is the range of the makespan itself? Equally, 
    how much do system factors cause the makespan to vary across identical ensemble
    workload executions?
\end{inparaenum}

Even if the makespan itself does not vary significantly across ensemble
workload executions, a large range of task runtimes within an ensemble
workload execution can waste computing resources, because the workload does
not complete until the last task has finished executing. Characterizing the
degree to which system factors affect individual task performance and the
makespan can help the understanding of these system (exogenous) factors (OS noise, inter-job interference, processor variability) and how they affect particular applications.
This study presents our observations on executing GROMACS ensemble workloads on the Theta cluster at the Argonne Leadership Computing Facility. GROMACS was chosen
partly because it is frequently executed as a part of ensemble workflows and
partly because of our familiarity with it.
In particular, we pay special attention to the 
\textit{methodology} employed to: (1) characterize the performance variation observed from executing large-scale 
GROMACS ensembles, and (2) systematically narrow down the list of the potential system (exogenous) factors 
responsible for the behavior.

\subsection{The Observation That We Seek to Characterize}
\label{sec:config}
\begin{figure}[htbp!]
 \vspace{-10pt}
  \centering
  \includegraphics[width=\columnwidth]{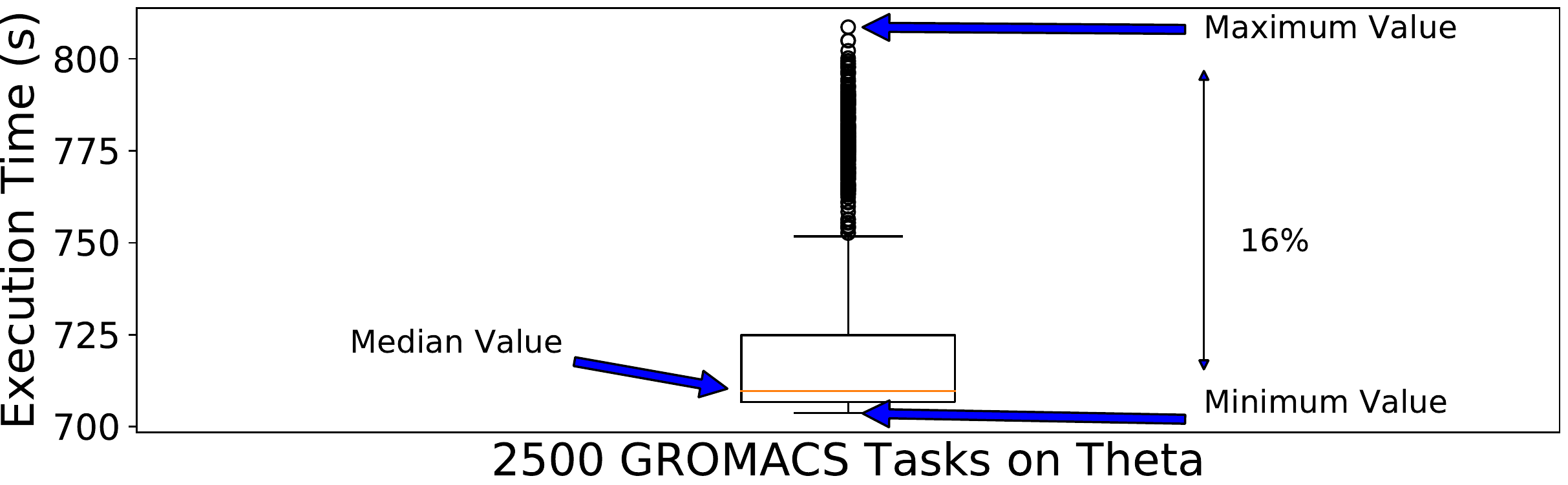}
  \caption{Performance Variation of Ensemble Tasks}
  \label{fig:behavior}
 \vspace{-10pt}
\end{figure}

Figure~\ref{fig:behavior} captures the behavior we seek to characterize and
investigate. A total of 2,500 GROMACS ensemble tasks belonging to ensemble
workloads spread across 10, 256-node batch jobs. Each of these GROMACS tasks
employs 64 MPI ranks and executes 100,000 simulation timesteps with OpenMP
support turned off. By default, checkpointing is enabled once every minute of
wallclock execution time. Further, these tasks operate on \textit{identical}
input decks and execute on \textit{identical} resource allocations (single,
64-core Theta KNL node). The task time-to-execution displays \textbf{16\%} of a difference between the
runtimes for the fastest and the slowest tasks. The experimental setup would
imply that the system (exogenous) factors are solely responsible for causing
the variation in task performance, requiring further investigation and root
cause analysis.

The observations in Figure~\ref{fig:behavior} provide an answer to the Question
1 when the tasks are spread across multiple ensemble workloads.
The question
of whether or not the number of batch jobs used to execute 2,500 tasks makes a difference to the result is indeed open. Further, when executing on a small scale (development queue, less than 8 KNL nodes), the makespan of the resulting ensemble workload executions displayed little to no variation. The runtime of the ensemble tasks within any given workload fell within tight upper and lower limits. A larger ensemble workload was required to \textit{consistently} observe any significant degree of performance variation.
Unless specified otherwise, the experimental results that follow assume the ensemble workload configuration described in Section~\ref{sec:config}.

\subsection{Characterizing Temporal Aspects of the Behavior}
\begin{figure}[htbp!]
 \vspace{-5pt}
\centering
  \includegraphics[width=\columnwidth]{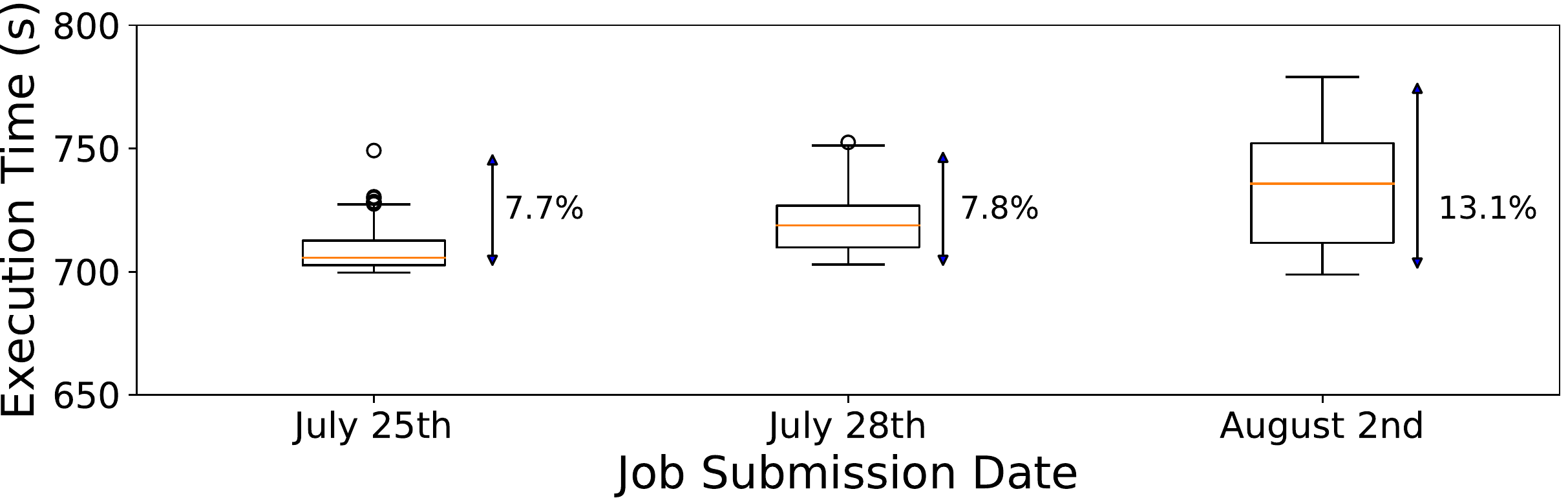}
 \caption{Variation in the Ensemble Makespan}
 \label{fig:temporal}
 \vspace{-10pt}
\end{figure}

We characterize the temporal aspects of the performance variation by investigating: (1) the range of individual task runtimes within a given ensemble workload, and (2) the range
of the makespan as a function of when the batch-job containing the ensemble workload executes. In the process, we wanted to identify if there was any correlation between the machine load (state) and the performance variations observed. We submitted several identical, 256-node GROMACS 
ensemble workloads for execution on different days and at different times during the day. Some of
these ensemble workloads were deliberately submitted during the night when the machine load
was low with few jobs actively executing. 
Figure~\ref{fig:temporal} depicts
the resulting plots of the task execution times from ensemble workloads that were executed
on three different days. These results indicate that the makespan varies significantly across time. 

The range of individual task runtimes within an ensemble workload also varies
across time. Importantly, we found no correlation between these
variations and the machine load. Therefore, we can safely disregard the job placement and network interference from
other jobs as likely factors.
The results from this experiment confirmed our
expectation, viz., each GROMACS ensemble task executed on a dedicated Theta
KNL node, effectively sandboxing the task from interfering or being interfered
with by other jobs running on the machine. The observations from this study
strongly suggested that whatever system factor was responsible for causing the
variation was either (1) local to each KNL node and was triggered
independently of the state of other nodes, or (2) common across all nodes.

\subsection{Ruling out ``Bad'' Nodes}
Our next set of experiments were designed to identify any persistent ``bad''
nodes in the system, i.e., nodes that \textit{consistently} worsened the
performance of any GROMACS ensemble task assigned to execute on them. Note
that we assume a general, hardware-based definition for what could cause a
node to be ``bad'', ranging from firmware issues to faulty power and CPU clock
frequency management. To test this ``bad node'' hypothesis, we set up the
experiment as follows:
\begin{inparaenum}
    \item Submit a 256-node batch job request;
    \item Run three identical, 256-task ensemble workloads successively on the same batch job allocation;
    \item Record the task ID, node ID, and the execution time of each task; and
    \item Correlate the execution time of each task and the node ID for the task offline.
\end{inparaenum}

\begin{figure}[htbp!]
 \vspace{-10pt}
\centering
  \includegraphics[width=\columnwidth]{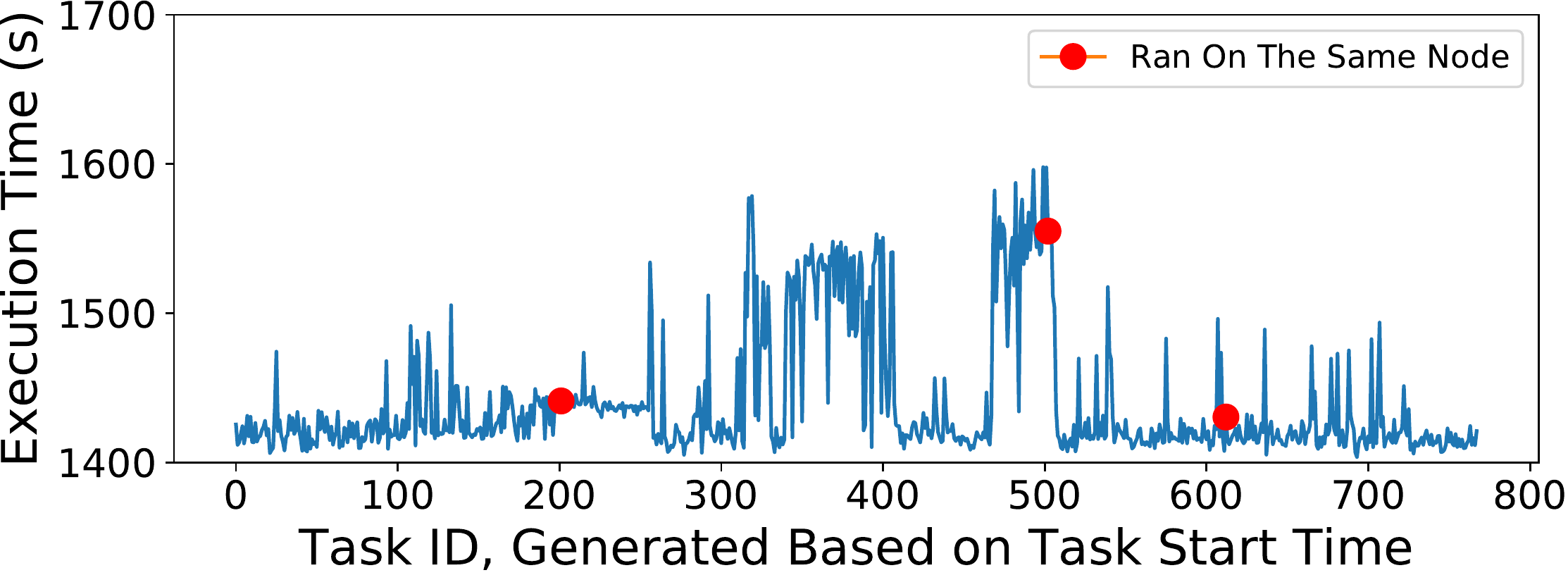}
 \caption{Testing The ``Bad Node'' Hypothesis}
 \label{fig:bad_nodes}
 \vspace{-10pt}
\end{figure}

In this experiment, each GROMACS task executed 200,000 timesteps of the main simulation loop, and checkpointing was enabled. Figure~\ref{fig:bad_nodes} depicts the execution time (Y-axis) as a function of task ID (X-axis). Tasks with IDs in the range 0-255 belong to the first ensemble workload, tasks with IDs in the range 256-511 belong to the second ensemble workload, and tasks with IDs in the range 512-755 belong to the third ensemble workload. The node ID
was generated based on the result of the \verb+hostname+ command. A careful analysis of the execution
times of GROMACS tasks and the nodes they were assigned to suggested that the ``bad node'' hypothesis did not hold water. As Figure~\ref{fig:bad_nodes} illustrates, tasks with IDs 201, 505, and 612 (marked as red dots)
ran on the same KNL node but displayed vastly different execution times. 
\subsection{Instrumentation to Understand the Affected Routines}
\begin{figure}[htbp!]
 \vspace{-5pt}
\centering
  \includegraphics[width=\columnwidth, keepaspectratio]{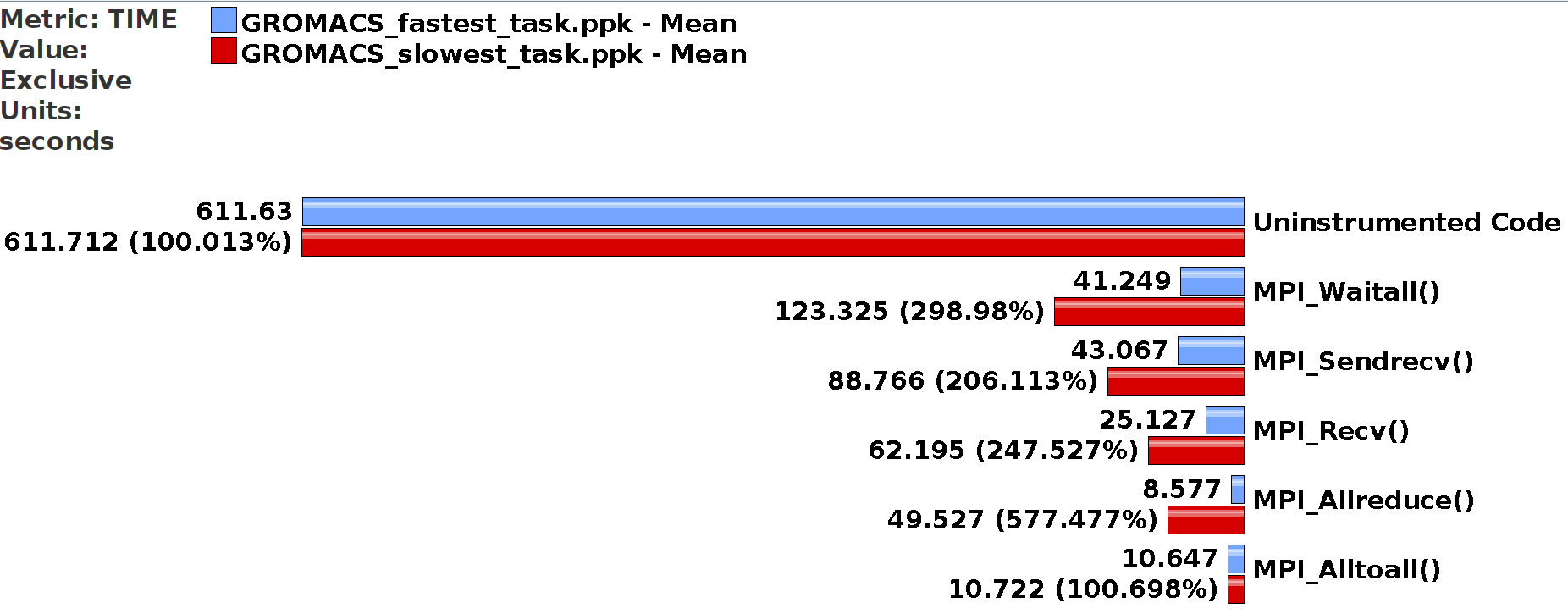}
 \caption{TAU Profiles for Fastest and Slowest Tasks}
 \label{fig:tau}
 \vspace{-10pt}
\end{figure}

With the ``bad node'' hypothesis effectively ruled out, we focused on identifying the GROMACS routines affected by the system factors. Specifically, through instrumentation, we wanted to account for
``extra'' time spent by the slowest task compared to the fastest.
The TAU Performance System~\cite{TAU} was employed 
to wrap MPI and POSIX I/O routines using lightweight library interposition techniques. Figure~\ref{fig:tau} is a ``diff'' or a comparison of two TAU profiles --- one for the
fastest GROMACS task and one for the slowest GROMACS task in an affected workload. To our surprise, we found that the ``extra'' time in the slowest task was attributed \textit{entirely} to four
routines --- \verb+MPI_Wait+, \verb+MPI_Allreduce+, \verb+MPI_Sendrecv+,
and \verb+MPI_Recv+. I/O routines were also affected but to a lesser extent. 

\subsection{Toggling Checkpoint and Parallel File I/O}
With the insight gained from profiling tasks with min-max execution times, our
focus turned to identify if the system factor was local to each node,  or not,
i.e., commonly present across multiple (or all) nodes. The network was
previously ruled out as a contributor, leaving the parallel filesystem as the
only candidate fitting the bill. Previous experiments enabled checkpointing at
a frequency of once a minute, and each GROMACS task produced a 2~MB
checkpoint file. Further, a deep dive into the GROMACS checkpoint
code module revealed that MPI rank 0 was responsible for performing the file
I/O after receiving data from other participating MPI ranks.

\begin{figure}[htbp!]
 \vspace{-5pt}
\centering
  \includegraphics[width=\columnwidth]{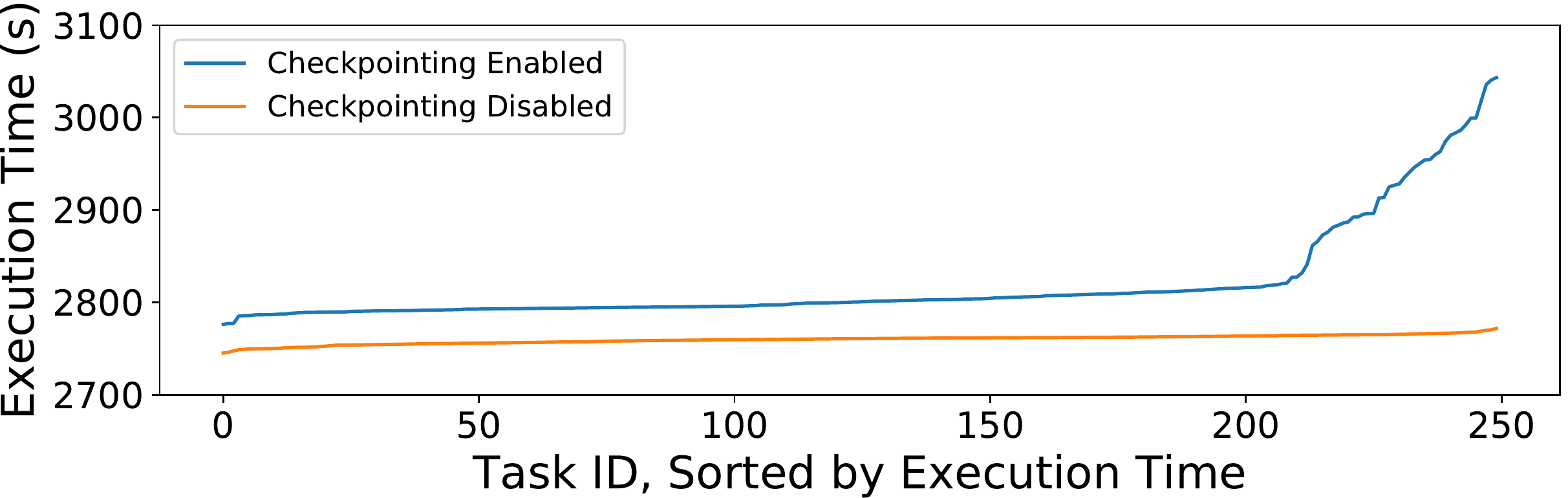}
 \caption{Runtime Variation with Checkpoint Enabled/Disabled}
 \label{fig:enabled_disabled}
 \vspace{-10pt}
\end{figure}

We set up an experiment to determine if disabling the checkpoint module would
result in greater consistency in GROMACS execution time. Two 256-node ensemble
workloads were simultaneously submitted, one with the checkpoint module
enabled and the other with the checkpoint module disabled. To ensure that the
tasks run for a sufficiently long time to allow the system factor a good
chance to ``show itself'' during the execution, the number of simulation
timesteps was set to 400,000. 
Figure~\ref{fig:enabled_disabled} depicts the \textit{sorted} task execution times of tasks in these two
ensemble workloads. The orange line represents the ensemble workload where checkpointing was disabled, and the blue line represents the workload where checkpointing was enabled. Checkpointing is indeed correlated with the occurrence of the variation in task runtime.

Next, we identified the line in the checkpoint module representing the file
\verb+write+ call. To identify the conditions necessary for the system factor
to show up, we requested a 256-node batch job allocation and ran four 256-task
ensemble workloads in the following configuration:

\begin{inparaenum}
    \item \textbf{YY}: Checkpoint enabled (\textbf{Y}) and file I/O enabled (\textbf{Y});

    \item \textbf{YN}: Checkpoint enabled (\textbf{Y}) and file I/O disabled (\textbf{N});

    \item \textbf{NY}: Checkpoint disabled (\textbf{N}) and file I/O enabled (\textbf{Y}); 

    \item \textbf{NN}: Checkpoint disabled (\textbf{N}) and file I/O disabled (\textbf{N}).
\end{inparaenum}

\begin{figure}[htbp!]
 \vspace{-10pt}
\centering
  \includegraphics[width=\columnwidth]{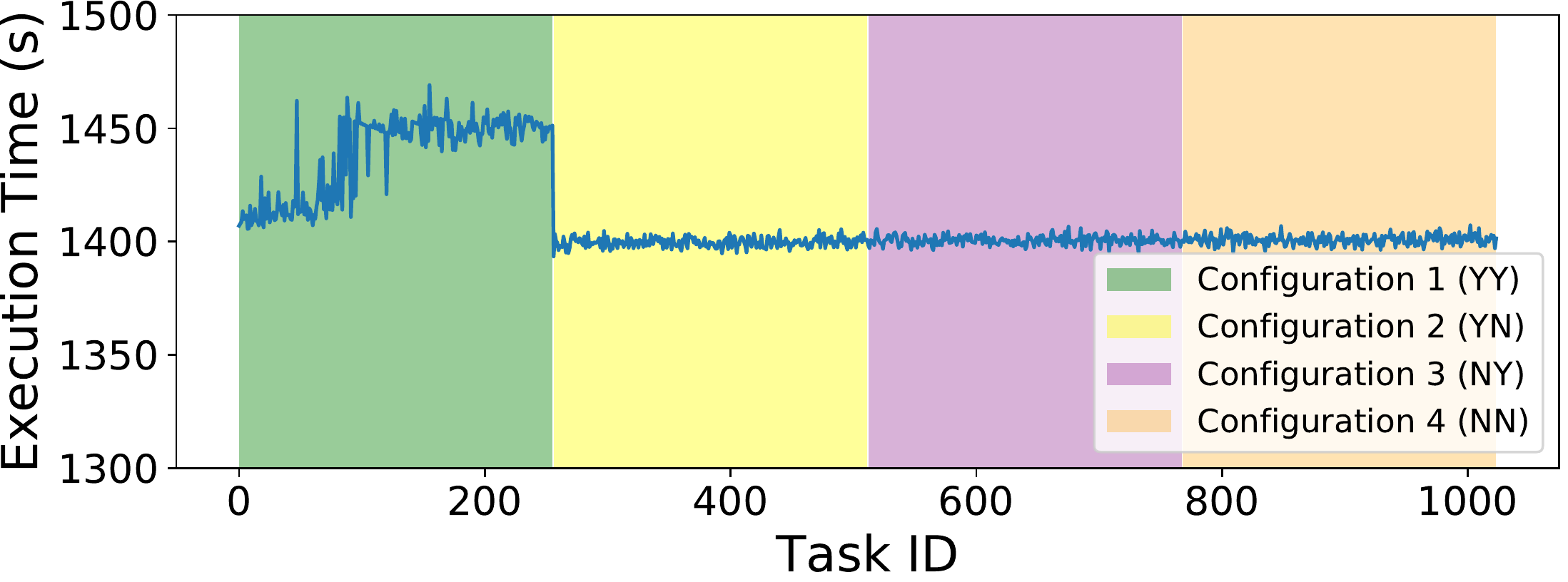}
 \caption{Toggling Checkpointing and File I/O}
 \label{fig:toggling}
 \vspace{-10pt}
\end{figure}

In Figure~\ref{fig:toggling}, tasks with IDs in the range 0-255 belong to
configuration 1, tasks with IDs in the range 256-511 belong to configuration
2, tasks with IDs in the range 512-767 belong to configuration 3, and tasks
with IDs in the range 768-1023 belong to configuration 4. System factors causing the runtime variation is triggered if checkpointing
is enabled \textit{and} the file I/O is enabled.

These results obtained from executing configuration 3 exonerate the parallel file system as the root cause
for the variation. If the parallel file system was indeed the root cause, then the ``diff'' in the routine
execution times (see Figure~\ref{fig:tau}) between the fastest and the slowest tasks would have shown up
in the \verb+write+ call. However, this is not the case --- instead, the ``diff'' in the routine execution
times shows up inside MPI routines. To further rule out the parallel file system as the cause,
we set up an experiment involving the IOR synthetic parallel file I/O benchmark. The benchmark
was configured to write 2 MB worth of ``dummy'' checkpoint data at a frequency of once every
minute, thereby emulating the GROMACS checkpoint module. Identical copies of the IOR benchmark
were executed inside an ensemble, and the ensemble workload was gradually scaled up from one to 256 KNL nodes on Theta. Regardless of the ensemble workload size, the IOR task performance remained
virtually constant (i.e., no runtime variation was observed between the IOR tasks).

\subsection{Inducing Performance Variation}
Assuming the system factor ``shows up'' uniformly for all applications, why is
GROMACS affected while the IOR benchmark was not? The experimental
observations thus far suggest that the MPI call structure surrounding the
GROMACS checkpoint module makes it \textit{particularly} sensitive to whatever
system factor was responsible for causing the variation. We wondered if the
``sensitivity'' of the checkpoint module execution time to minor variations in
the file \verb+write+ time can be triggered by replacing the \verb+write+ call
with a \verb+sleep+ call for a ``random'' amount of time in the range of 0.09
to 0.25 seconds. These numbers were chosen based on the time to write a 2 MB
checkpoint file.

\begin{figure}[htbp!]
 \vspace{-10pt}
\centering
  \includegraphics[width=\columnwidth]{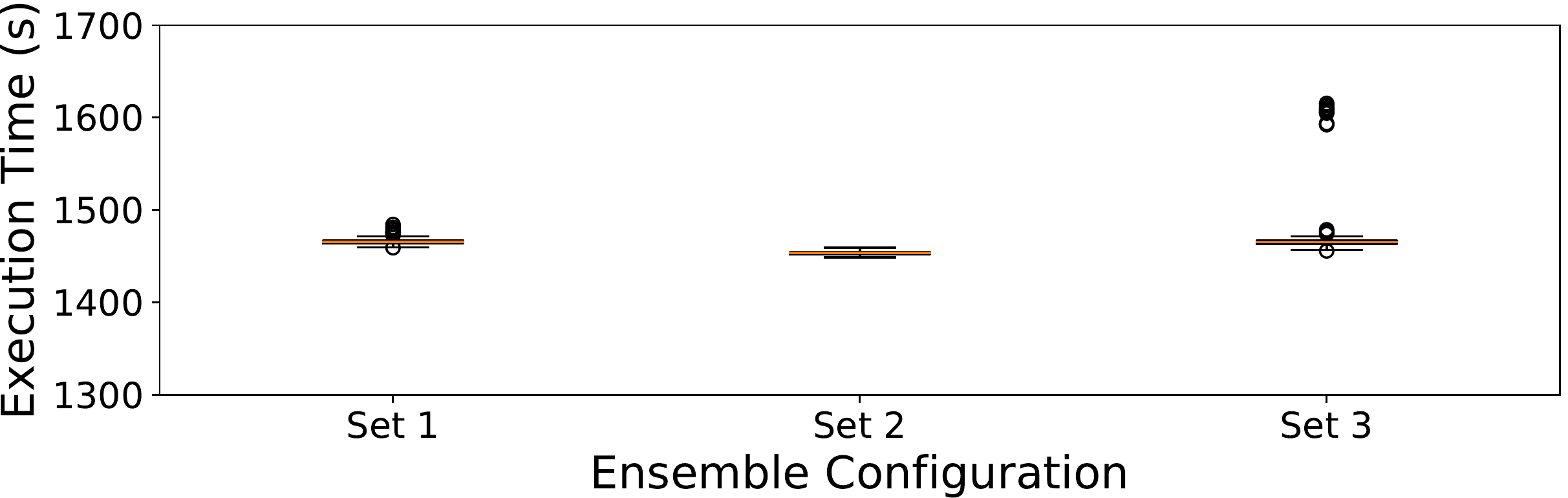}
 \caption{Induced Performance Variation}
 \label{fig:induced}
 \vspace{-15pt}
\end{figure}

We ran three successive sets of 256-node GROMACS ensembles within the same batch job. For Set 1, the checkpoint module and file I/O were
enabled. For Set 2, the checkpoint module and file I/O were disabled, and for Set 3, the checkpoint 
module was enabled, but the file \verb+write+ call was replaced with a random \verb+sleep+ call. 
Figure~\ref{fig:induced} depicts the results of this experiment. Set 1 displays a moderate degree of
variation in the task runtime while Set 2 displays a minimal degree of variation (expected). 
Surprisingly, Set 3, which replaces the file \verb+write+ call with an ``equivalent'' \verb+sleep+ call
displays the highest degree of variation in task runtime. Specifically,
the \verb+sleep+ call perturbed the execution of a few outlier tasks to the extent of adding \textit{hundreds} of seconds to the total execution time. 
\subsection{Relying on Prior Work}
In summary, the observations from our experiments suggest that:
\begin{inparaenum}
    \item The system factor causing the variation is \textit{local} to each KNL node (and task);
    \item Its occurrence and the severity of the impact
    on task execution time can not be predicted
    but is bounded;
    \item The system factor is \textit{not} the
    result of faulty hardware; and
    \item GROMACS performance is susceptible to the
    presence of this system factor when checkpointing
    is enabled.
\end{inparaenum}    

The \verb+sleep+ call and the \verb+write+ call have a common ``pathway'' or element causing the behavior we seek to understand. Our collective hypothesis at this point was 
that the system factor is OS noise, but we did not know how to test this hypothesis.
A brief review of the existing literature on prior performance variation studies yielded valuable insight. Chunduri et al.~\cite{CHUNDURI} report on the exogenous system factors known to cause performance variation on the Theta cluster.
As part of their study, they investigated the impact of OS noise on the execution time of the
Selfish benchmark~\cite{SELFISH} with and without ``core specialization'' ---
a mechanism allowing the user to dedicate a set of cores to execute OS services.
They observed a ten-fold reduction in noise when core specialization was enabled. However, they did not explore the impact of OS noise
on production HPC applications.

\begin{figure}[htbp!]
 \vspace{-5pt}
\centering
  \includegraphics[width=\columnwidth]{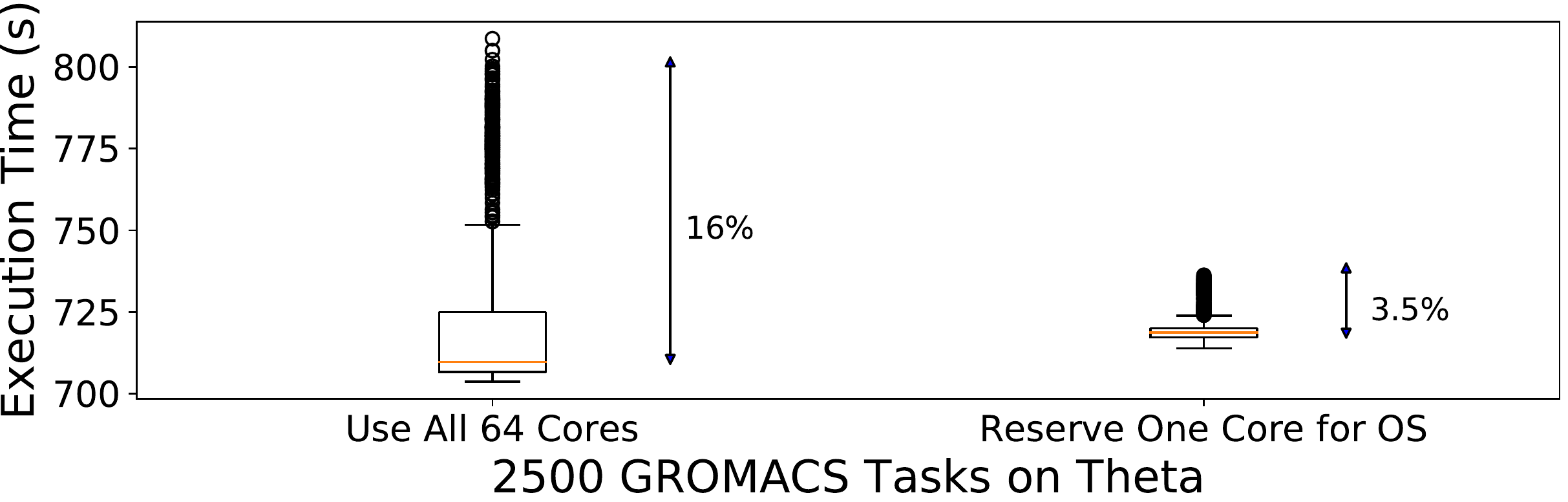}
 \caption{Reserving a Dedicated Core for OS Services}
 \label{fig:reserve}
 \vspace{-15pt}
\end{figure}

By default, core specialization on Theta is disabled, allowing the OS to select any core to schedule
its services. 
We modified the GROMACS task configuration to execute on 63 KNL cores instead of using all the 64 cores, reserving one core on every node for the OS to schedule its services. Over several days, we executed
ten 256-node ensemble workloads in this new configuration. Figure~\ref{fig:reserve}
depicts the results of employing a dedicated core for scheduling OS services --- the range of GROMACS task execution time reduced
from 16\% (when using all 64 cores) to 3.5\% (when using 63 cores).
As a result of using one less CPU core, the individual task performance \textit{worsens} by 1-2\% but the \textit{collective} performance of
the workload improves.

\section{Discussion}

Although we reduced task runtime variation putatively caused by OS noise, we
still have no good way of proving through a set of measurements that the OS
noise was indeed the dominant cause.  By running all MPI processes of a
GROMACS task on a single Theta node, network contention between the processes
is minimized (all communication is through shared memory), ruling this out as
a cause for the differential timings in the MPI routines.  Is there another
easily measured metric that might correlate to OS noise?  
The number of
process context switches is a possible candidate, since it might suggest more
direct OS activity, but it is more likely what happens between process context
switch that matters rather than the number.  We might try to measure the time
spent in the OS, but there might be little variation, assuming the OS is doing
similar tasks on each node.  What is weird is that only a percentage of tasks
are affected and the variation is showing up in MPI routines.  This suggests
that it matters \textit{where} in the task's
execution the OS interference occurs,
not how much time it lasts.
We speculate that some of the GROMACS tasks are
just unlucky and the OS interfers at times that
disrupts MPI behavior, adding delays and impacting
syncronication.

There is a need to fundamentally understand the genesis of performance
variation resulting from exogenous and endogenous factors and study their
impact on critical HPC applications such as GROMACS.  Prior studies have been
primarily conducted on (micro)benchmarks; it is largely unclear how these
results translate to production HPC applications.  Having some systematic
means to perturb application execution might be useful to explore performance
sensitivities. However, modeling the impact of task level performance
variation on HPC ensembles is a different endeavor altogether.  A more
hierarchical methodology might be required: performance variation observed at
a task level may, or may not, translate to performance variation at an ensemble
workload level, especially if there are several more tasks to execute than
available computing resources.  


Before discussing performance reproducibility in ensemble computations, it is
beneficial to reflect on performance variation assumptions of an HPC
application. It is generally expected that a deterministic application code
will exhibit minor, if any, relevant variation in its performance if it is run
in the same configuration on the same machine. It is this expectation that
forms the basis for HPC performance optimization where inefficiencies can be
reliably identified and improvements made. That said, several exogenous
factors can lead to performance variation, including differences in processor
clock rates, contention with other running codes, operating system
interference, and more. Endogenous to an application, there can be
non-determinism inherent in its execution and fine sensitivities to memory and
network operation.

Nevertheless, it is reasonable to maintain the expectation of
minor performance variation in HPC applications; otherwise, our
performance optimization methodology breaks down. It might be
even assumed possible to characterize any variability within the
context of the application code, in order to determine bounds
and vulnerabilities. With this in mind, performance variation
for an HPC application can be generally regarded as a distribution
around expected performance value, one that very well might be
able to be reproduced. 

However, ensemble execution is a different matter. All factors for performance
variation of a single HPC application are still valid for ensemble tasks.
Instead of a single code, the variability analysis is compounded by the
multiplicity of tasks.
Indeed, tasks can interfere with each other as a
result of sharing system infrastructure and cause
perturbations that lead to further performance variation.
In addition, it is essential to realize that how a
task is configured at the time of its execution and the resources assigned are not necessarily fixed.
The same task might be run in a variety of configurations during the lifetime
of the ensemble computation. It is not the optimization of an individual task
under a particular configuration that matters. Task performance naturally has
a much fuller distribution because it covers a potentially broader set of
options that the ensemble runtime system has to choose from during task
scheduling.

A fundamental shift arising from ensemble (task) performance variability is
that it is no longer the performance or optimization of a single task that is
critical, but the collective performance of all ensemble members (viz.,  the
makespan of the workload that matters).
What implications does this have on
performance reproducibility in ensemble computations?  Are we asking for
individual task performance to be reproducible, and by extrapolation, thinking
the ensemble's performance should then be reproducible? Given that an ensemble
workflow maintains task dependencies, does this imply that reproducibility
will require invariance in task execution ordering from one ensemble run to
the next?


We contend that the focus of reproducibility should not be on whether an
individual task's performance is reproducible in ensemble computation. When a
task is ready to run, it will then be scheduled and configured according to
the resources allocated. If the task is run again at another point in the
workflow, it might be configured differently and with different resources. It
makes little sense to compare the two task executions if they are configured
differently -- a 4-node MPI+OpenMP task with 32 CPU threads per node will
behave very differently from a 1-process instantiation of the task with 4
GPUs. Here performance variation stares the ensemble scheduler directly in the
face. Furthermore, decisions regarding resource assignment might have less to
do with optimal task performance than with dynamic resource availability and
collective utilization of ensemble resources. Even as we relinquish the belief
of single task performance reproducibility, we are constrained by need for
collective performance invariance (viz., makespan) and reproducibility.

It is worthwhile to define three regimes of ensemble execution
relative to resources and workload (tasks ready to run):
(1) there are inadequate resources to run the entire workload,
(2) there are adequate resources available to execute
the workload, and
(3) there are excessive (more than adequate) resources available
to execute the workload.
Understanding that the workload amount at any point in the
ensemble computation will vary, without loss of generality,
we can characterize the regimes.
In (1), the appropriate and optimal subset of tasks
will have to be selected and ordered to complete the workload.
That decision is non-trivial and can involve other criteria
than task performance per se. While tasks in a workload set
are independent, some might be determined to be of higher
priority.
In (3), the correct subset of
resources have to be selected, and the rest possibly released
to save cost.
In (2), the mapping of tasks to resources (which assumes
alternative task configurations) has to be optimal.
In all regimes, we are dealing with reasoning about the
workload performance concerning the collective.

Even if we knew the expected performance of each
task instance (for possible configurations and resource assignment), the scheduling
problem is difficult \textit{a priori}.
Allowing for even a tiny
variation in any task's performance makes finding the
``optimal'' schedule significantly harder. A distribution
of solutions will very likely be required.
Knowing whether a task's execution will be reproducible does little to make
the problem tractable.
If we forget about whether
task performance is reproducible, how should we define ``reproducibility'' in
the context of ensembles? One possible way to approach this is to simplify the
focus of the problem of scheduling the \textit{next} task. Indeed, given an
executing subset of workload tasks, the ensemble runtime system must
a) decide which task to execute next, b) select/allocate free
resources to that task, and c) configure the task appropriately.
Thus, the problem is reduced to deciding on the next task
and its execution.

It is not to say that the problem is now easy. If we had any
information about the performance of the remaining tasks,
that might be beneficial, but we have to assume it
will have performance variability. We may also need to consider non-stationary distributions from which individual task performance is drawn.
Having no knowledge of expected performance is
equivalent to high variability. Once a task begins, its
performance is determined by its execution. The workload performance as a collective will be determined by the
\textit{next} task assignment, task by task.
Interestingly, at those points extra information will
be available about which task(s) finished and how
resource availability may have changed as a result.
Furthermore, it is possible to gain further information
and insight into how well individual tasks are performing
by employing support for observation and monitoring
in the ensemble system.

In one sense, then, we could reason about ensemble reproducibility concerning
whether the same decision would be made about scheduling the next task if the
circumstances were the same in different ensemble executions, for every
ensemble task. Unfortunately, this reasoning falls on its face since any task
performance variation has the potential to make each ensemble execution
different. If, instead, we recast reproducibility as making ``locally
optimal''scheduling decisions, there is more consistency across multiple runs
of an ensemble computation that we can imagine.

Carrying this further, if we rationalize ensemble
computing concerning its potential for massive
makespan, we could imagine that its performance will
begin to conform to the \textit{Law of Large Numbers}.
That is, the greater the ``span'' of tasks off the critical
path, the less impact variations among those tasks will have
acute effects, even if they are scheduled differently.
That is rationally sound but likely difficult to actually
prove experimentally.
Lastly, although we have downplayed adaptability in this
paper, the institution of an observation and monitoring
framework would allow not only increased performance awareness
to better inform future scheduling decisions, but also
enable detection of performance anomalies when they arise,
allowing corrective actions, however possible. The situational
awareness of performance state coupled with in situ analytics
and dynamic response to problematic behavior will make
observation, monitoring, and adaptive control critical for
harnessing reproducibility in ensemble computing.

\footnotesize{This work was partially supported by the DOE HEP Center for
Computational Excellence at Brookhaven National Laboratory under B\&R
KA2401045, as well as NSF-1931512 (RADICAL-Cybertools). We thank Andre Merzky and Li Tan for useful discussions and support.}

\bibliographystyle{IEEETran}
\bibliography{radical,sigproc}

\begin{thebibliography}{10}
\providecommand{\url}[1]{#1}
\csname url@samestyle\endcsname
\providecommand{\newblock}{\relax}
\providecommand{\bibinfo}[2]{#2}
\providecommand{\BIBentrySTDinterwordspacing}{\spaceskip=0pt\relax}
\providecommand{\BIBentryALTinterwordstretchfactor}{4}
\providecommand{\BIBentryALTinterwordspacing}{\spaceskip=\fontdimen2\font plus
\BIBentryALTinterwordstretchfactor\fontdimen3\font minus
  \fontdimen4\font\relax}
\providecommand{\BIBforeignlanguage}[2]{{%
\expandafter\ifx\csname l@#1\endcsname\relax
\typeout{** WARNING: IEEEtran.bst: No hyphenation pattern has been}%
\typeout{** loaded for the language `#1'. Using the pattern for}%
\typeout{** the default language instead.}%
\else
\language=\csname l@#1\endcsname
\fi
#2}}
\providecommand{\BIBdecl}{\relax}
\BIBdecl

\bibitem{jha2019incorporating}
S.~Jha, S.~Lathrop \emph{et~al.}, ``{Incorporating Scientific Workflows in
  Computing Research Processes},'' \emph{Computing in Science \& Engineering},
  vol.~21, no.~4, pp. 4--6, 2019.

\bibitem{lee2020scalable}
H.~Lee, A.~Merzky \emph{et~al.}, ``Scalable hpc and ai infrastructure for
  covid-19 therapeutics,'' \emph{Platform for Advanced Scientific Computing
  Conference (PASC ’21), July 5–9, 2021, Geneva, Switzerland. ACM, New
  York, NY, USA}, 2021, \url{https://arxiv.org/abs/2010.10517}.

\bibitem{brace2022coupling}
A.~Brace, I.~Yakushin \emph{et~al.}, ``Coupling streaming ai and hpc ensembles
  to achieve 100--1000$\times$ faster biomolecular simulations,'' in \emph{2022
  IEEE International Parallel and Distributed Processing Symposium
  (IPDPS)}.\hskip 1em plus 0.5em minus 0.4em\relax IEEE, 2022, pp. 806--816.

\bibitem{balasubramanian2018harnessing}
V.~Balasubramanian, M.~Turilli \emph{et~al.}, ``Harnessing the power of many:
  Extensible toolkit for scalable ensemble applications,'' in
  \emph{International Parallel and Distributed Processing Symposium}.\hskip 1em
  plus 0.5em minus 0.4em\relax IEEE, 2018, pp. 536--545.

\bibitem{cosb18kasson}
P.~M. Kasson and S.~Jha, ``Adaptive ensemble simulations of biomolecules,''
  \emph{Current Opinion in Structural Biology}, vol.~52, pp. 87 -- 94, 2018.

\bibitem{balasubramanian2020adaptive}
V.~Balasubramanian, T.~Jensen \emph{et~al.}, ``Adaptive ensemble biomolecular
  applications at scale,'' \emph{SN Computer Science}, vol.~1, no.~2, pp.
  1--15, 2020, \url{https://doi.org/10.1007/s42979-020-0081-1}.

\bibitem{merzky2022raptor}
A.~Merzky, M.~Turilli, and S.~Jha, ``Raptor: Ravenous throughput computing,''
  in \emph{2022 22nd International Symposium on Cluster, Cloud and Internet
  Computing (CCGrid)}, 2022, pp. 595--604.

\bibitem{pouchard2019computational}
L.~Pouchard, S.~Baldwin \emph{et~al.}, ``Computational reproducibility of
  scientific workflows at extreme scales,'' \emph{The Int. Journal of High
  Performance Computing Applications}, vol.~33, no.~5, pp. 763--776, 2019.

\bibitem{pouchard2018use}
------, ``Use cases of computational reproducibility for scientific workflows
  at exascale,'' \emph{arXiv preprint arXiv:1805.00967}, 2018.

\bibitem{PROCESSOR}
Y.~Inadomi, T.~Patki \emph{et~al.}, ``Analyzing and mitigating the impact of
  manufacturing variability in power-constrained supercomputing,'' in
  \emph{SC'15: Proceedings of the International Conference for High Performance
  Computing, Networking, Storage \& Analysis}.\hskip 1em plus 0.5em minus
  0.4em\relax IEEE, 2015, pp. 1--12.

\bibitem{NETWORK}
A.~Bhatele, K.~Mohror \emph{et~al.}, ``There goes the neighborhood: performance
  degradation due to nearby jobs,'' in \emph{SC'13: Proceedings of the
  International Conference on High Performance Computing, Networking, Storage
  and Analysis}.\hskip 1em plus 0.5em minus 0.4em\relax IEEE, 2013, pp. 1--12.

\bibitem{CHUNDURI}
S.~Chunduri, K.~Harms \emph{et~al.}, ``Run-to-run variability on xeon phi based
  cray xc systems,'' in \emph{International Conference for High Performance
  Computing, Networking, Storage \& Analysis}, 2017, pp. 1--13.

\bibitem{TAU}
S.~Shende and A.~Malony, ``The {TAU} parallel performance system,'' \emph{The
  International Journal of High Performance Computing Applications}, vol.~20,
  no.~2, pp. 287--311, 2006.

\bibitem{SELFISH}
P.~Beckman, K.~Iskra \emph{et~al.}, ``Benchmarking the effects of operating
  system interference on extreme-scale parallel machines,'' \emph{Cluster
  Computing}, vol.~11, no.~1, pp. 3--16, 2008.

\end{thebibliography}
\end{document}